\documentclass[preprint]{aastex}

\shorttitle{$\kappa^1$~Ceti}
\shortauthors{Rucinski \& et~al.}

\begin{document}

\title{Differential rotation of the active G5V star
$\mathbf{\kappa^1}$ Ceti: Photometry from 
the MOST satellite\footnote{Based 
on data obtained with the MOST satellite: a Canadian Space Agency 
mission, jointly operated by Dynacon Inc., 
the University of Toronto Institute of 
Aerospace Studies and the University of British Columbia.
}}

\author{Slavek M. Rucinski\altaffilmark{2}}
\affil{David Dunlap Observatory, University of Toronto \\
P.O.~Box 360, Richmond Hill, ON L4C 4Y6, Canada}
\email{rucinski@astro.utoronto.ca}

\author{Gordon A.H. Walker\altaffilmark{2}, 
Jaymie M. Matthews\altaffilmark{3}, Rainer Kuschnig\altaffilmark{4}, 
Evgenya Shkolnik\altaffilmark{2}}
\affil{Department of Physics and Astronomy, University of British Columbia \\ 
6224 Agricultural Road, Vancouver, BC V6T 1Z1, Canada}
\email{walker@uvic.ca, (matthews,kuschnig,shkolnik)@astro.ubc.ca}

\altaffiltext{2}{Visiting Astronomer, Canada-France-Hawaii Telescope,
operated by the National Research Council of Canada, the Centre
National de la Recherche Scientifique of France, and the University of
Hawaii.}

\altaffiltext{3}{MOST Mission Scientist}

\altaffiltext{4}{MOST Instrument Scientist}

\author {Sergey Marchenko}
\affil{Department of Physics and Astronomy, 
Western Kentucky University\\ 1 Big Red Way, Bowling Green, KY 42101}
\email{sergey.marchenko@wku.edu}

\author {David A. Bohlender\altaffilmark{2}} \affil{National Research
Council of Canada, Herzberg Institute of Astrophysics\\ 5071 West
Saanich Road Victoria, BC, Canada V9E 2E7}
\email{david.bohlender@nrc.ca}

\author{D. B. Guenther}
\affil{Department of Astronomy and Physics, St. Mary's University\\
Halifax, NS B3H 3C3, Canada}
\email{guenther@ap.stmarys.ca}

\author{Anthony F.J. Moffat}
\affil{D\'epartement de physique, Universit\'e de Montr\'eal \\ 
C.P.\ 6128, Succ.\ Centre-Ville, Montr\'eal, QC H3C 3J7, Canada}
\email{moffat@astro.umontreal.ca}

\author{Dimitar Sasselov}
\affil{Harvard-Smithsonian Center for Astrophysics \\ 
60 Garden Street, Cambridge, MA 02138, USA}
\email{sasselov@cfa.harvard.edu}

\author{Werner W. Weiss}
\affil{Institut f\"ur Astronomie, Universit\"at Wien \\ 
T\"urkenschanzstrasse 17, A--1180 Wien, Austria}
\email{weiss@astro.univie.ac.at}

\begin{abstract}
About 30.5 days of nearly uninterrupted broadband photometry of the 
solar-type star $\kappa^1$~Ceti, obtained with the MOST 
(Microvariability \& Oscillations of STars) satellite, shows 
evidence for two large starspots with different rotation periods
of 8.9 and approximately 9.3 days ($\Delta\Omega / \Omega \simeq 4\%$).  
Ground based measurements in 2002 and 2003 of Ca II H \& K 
emission reveal variations in chromospheric activity with 
a period of about 9.3 days. The data were obtained during the MOST 
commissioning phase. When the data are combined with 
historical observations, they indicate that the 9.3-day
spot has been stable in its period for over 30 years.  
The photometry, with a sampling rate of approximately 
once per minute, was also used to search for acoustic 
(p-mode) oscillations in the star. We detect no clear evidence for 
p-modes in the $\kappa^1$~Ceti photometry, with a noise level around 
7 -- 9 $\mu$mag at frequencies in the range 0.5 -- 4 mHz 
(3--$\sigma$ detection limit of 21 -- 27 $\mu$mag).  
There were no flares 
or planetary transits during the 30.5 days of MOST monitoring 
with light amplitudes greater than 2 mmag 
(durations greater than 200 minutes) and 3 mmag ($2 - 200$ 
min durations). While this rules out any close-in planets of 
$\geq$0.5 Jupiter diameters with an orbital 
inclination close to $90^{\circ}$, the scatter in differential 
radial velocities permit a close giant planet in a more highly 
inclined orbit.
\end{abstract}


\section{INTRODUCTION}
\label{intro}

The production of solar flares, the evolution and migration of sunspots, 
and the very origin of the Sun's magnetic field are all believed to be 
associated with differential rotation beneath and at the solar surface.  
To test models of stellar dynamos, measurements of differential 
rotation in other stars are necessary for correlation with other 
parameters like magnetic variability and chromospheric 
activity.  

Surface differential rotation (SDR) can be easily observed 
in the Sun -- with detailed 
observations going back to \citet{Carrington1860} -- 
and subsurface rotation has been inferred from helioseismic data 
of the Sun's 5-minute p-mode oscillations (e.g., \citet{GONG1996}).  
However, it is a challenge to observe SDR
directly at the surfaces of other stars. 
With the notable exception of EK~Dra, where
two spot modulations differing in period by 5.5\% are simultaneously 
visible \citep{MG2003}, evidence for SDR in most
solar-type stars with prominent starspots is 
seen only by comparing data over many years.  
Light variations ascribed to rotational modulation sometimes change 
in period from epoch to epoch, and the most natural explanation 
is that the dominant spot systems appear at 
different latitudes and move with different rotation periods 
(e.g., \citet{HB1990}; \citet{HEHH1995}). 
Doppler Imaging of some solar-type stars from epoch to epoch 
has also revealed evidence for SDR (see reviews by 
\citet{CollierCameron2002} and \citet{strass2004}) 
and in at least one case, the rapidly rotating K dwarf AB~Doradus, 
SDR was observed in the Doppler maps in only a few 
consecutive rotations (\citet{CCDS2002}). So far, 
there have not yet been measurements of solar-type acoustic oscillations 
in any star other than the Sun with sufficient frequency resolution 
to explore its interior rotation. 

The MOST (Microvariability \& Oscillations of STars) satellite 
\citep{MOST} was pointed at $\kappa^1$~Ceti as part of the 
commissioning phase of the mission, to obtain useful science 
during engineering tests and debugging of the MOST systems.  
Our space-based photometry has sufficient time coverage and precision to 
reveal SDR which is easily evident even in only 
three equatorial rotations of the star. The data also set the 
first meaningful limits on the amplitudes of p-mode
oscillations in this star.

$\kappa^1$~Ceti was chosen because of (1) its brightness and location 
in the sky at the time of these tests, (2) previous indications 
of spot migration and hyperactivity, and (3) its G5V spectral 
type, which made it a possible candidate for solar-type 
oscillations; we describe the star more fully in Section~\ref{kappa}.
The MOST observations are discussed in Section~\ref{obs}. 
The MOST-orbit binned data produce a light curve which reveals the 
flux rotational modulation of a young active Sun-like star in 
unprecedented completeness and precision (Section~\ref{lc}).
We are able to relate the photometric modulations of spot
visibility to periodic rotational variations of 
chromospheric activity seen in 
high-resolution spectroscopic observations of 
the Ca~II K emission in 2002 and 2003 (Section~\ref{CaII}), 
the latter obtained just before the MOST observations; the same 
data provided a new accurate determination of $V \sin i$ for
the star. The full temporal resolution MOST photometry 
is used to set meaningful limits on the 
oscillation amplitudes in this star in Section~\ref{p-mode}.  

\setcounter{footnote}{4}      

\section{THE STAR $\kappa^1$~Ceti}
\label{kappa}

$\kappa^1$~Cet (HD~20630, HIP~15457, HR~996; $V=4.83$, $B-V=0.68$)
is a nearby ($9.16 \pm 0.06$ pc, \citet{hip}) G5V dwarf. Its variability
with a period of 9.09 days was detected by the {\it Hipparcos\/} 
mission \citep{hip}. Since then, several studies were aimed at reconciling
apparent changes in the period, changes which can be explained
by different latitudes of spot formation in different years. 

As a MK spectral classification standard \citep{Gray2001},
$\kappa^1$~Ceti is one of the most frequently observed stars.  Although
sometimes considered a ``Very Strong Lined'' late-type dwarf, its
metallicity may be only slightly higher than solar, $[Fe/H]=+0.05 \pm
0.05$ (for full references, see \citet{HL2003}).  As far as it has been
possible to establish, it is a single star and does not possess any
large planets \citep{Halb2003}.  Its radial velocity of +18.9
km~s$^{-1}$ combined with the {\it Hipparcos\/} proper motions leads to
a rather moderate spatial velocity relative to the Sun,
suggesting Young Disk population membership (the two
available estimates disagree slightly: \citet{Gaid2000}, 
$U=-12.7$, $V=+7.1$ and $W=+2.6$;  \citet{Mont2001}
$U=-22.4$, $V=-4.3$ and $W=-5.3$).

\citet{Gud97} estimated an age of 750 Myr from the relatively 
rapid rotation of 9.2 days seen in the spot modulation 
and suggested that the star is a likely member of the Hyades moving
group. However, \citet{Mont2001} considered seven moving groups of
nearby young stars, but were unable to associate $\kappa^1$~Ceti 
with any of them.

The young age of the star was the reason for \citet{DG1994} to include
$\kappa^1$~Ceti in the {\it The Sun In Time\/} project which attempts to
characterize young solar-type stars in terms of temporal changes taking
place particularly at the epochs before terrestrial life formation
\citep{Guin2003}.  In the group of 6 such stars, with the youngest
being 130 Myr old, $\kappa^1$~Ceti is one of the most advanced ones with
an estimated age in the \citet{DG1994} study of about 650 Myr. The 
difference in the age of 100 Myr versus 750 Myr of \citet{Gud97} 
can be traced to the current uncertainty of estimates based on the 
rotation rate. (Note: The 8.9 day rotation period found in this paper 
may be taken as an indication of an even younger age). 

As is observed for stars with activity somewhat moderated with age, the 
star shows an activity cycle. \citet{Baliu1995} monitored the narrow-band 
Ca II H \& K chromospheric fluxes photoelectrically from 1967 to 1991 
expressing them in terms of the $S_{HK}$ index. They found a rotational 
period of $9.4 \pm 0.1$ days \citep{Bal83} with a chromospheric activity
cycle of 5.6 years \citep{Baliu1995}; the quality of the latter 
determination was described as ``fair'' and
longer term trends were noted. Using broad band photometry, 
\citet{MG2002} observed a photometric activity cycle of
of $5.9 \pm 0.2$ years between 1990 and 1999, with
a rotational period of 9.214 days.

On 24 January 1986, \citet{Rob88} caught the signature of a massive
flare in He~I~D$_{3}$ which \citet{Sch00} subsequently estimated 
had an energy $\sim 10^{35}$ ergs. \citet{Sch00} speculated that 
such massive flares on a
single solar-type star might be triggered by the magnetic field of a
close giant planet such as those in 51~Peg-type systems.  Neither 
\citet{Wal95} nor \citet{Cumming1999} detected any long-term periodic 
variability over many years
in their precise radial velocity (PRV) measurements, so there exist no 
dynamical perturbations consistent with a close-in giant planet  
(unless the orbit is highly inclined). On the other hand, 
\citet{Wal95} did find (their Figure~2) 
a rapid RV change of +80 m~s$^{-1}$ in 1988.5
which was accompanied by an equally rapid increase in chromospheric
activity in the Ca~II $\lambda$8662 line. While the RV change could be
modeled by the close approach of a giant planet in a highly elliptical
orbit, the tight correlation of the RV with chromospheric activity
(both positive and negative RV excursions) pointed to a change
intrinsic to the star. In Section~\ref{RV} we point out that, within the 
scatter of existing RV measurements, $\kappa^1$~Ceti could harbor 
in a close, high inclination orbit.

\section{THE MOST PHOTOMETRY}
\label{obs}

\subsection{Photometry of $\kappa^1$~Ceti}
\label{phot}

The photometric observations discussed in this paper were obtained by 
the MOST (Microvariability \& Oscillations of STars) 
microsatellite \citep{MOST} between November 5 and December 5, 2003,
during the commissioning phases of the mission (constraints 
and limitations of these phases are
described in the next Section~\ref{comm}).
MOST is Canada's first orbiting space telescope, launched in June 2003 to study 
(1) acoustic oscillations in solar-type and magnetic peculiar stars, (2) rapid 
variability in hot massive stars; and (3) reflected light from close-in giant 
exoplanets. The MOST instrument is an optical telescope (aperture 15 cm) feeding 
a CCD photometer through a single broadband optical filter (350 -- 700 nm).  
Thanks to its polar Sun-synchronous orbit (altitude 820 km), MOST can monitor 
stars in a zodiacal band of sky about $54^\circ$ wide for up to 8 weeks without 
interruption. Starlight from Primary Science Targets, such
as $\kappa^1$~Ceti, i.e.\ brighter than $V \simeq 6$, 
is projected onto the CCD as a fixed image of the telescope pupil covering about 
1500 pixels for high photometric stability and insensitivity to detector 
flatfield and radiation effects on individual pixels. As has already been 
demonstrated for Procyon \citep{Procyon}, MOST is capable of reaching 
in that mode a photometric precision of about $\pm 1$ part 
per million (ppm) to search for oscillations in very bright stars.

Despite limitations during commissioning (see the next section), 
the MOST data of $\kappa^1$~Ceti achieved a duty cycle of 
about 96\% over a time span of 30.5 days 
(with only three gaps of a few hours each), with 40-second 
exposures obtained approximately once per minute.
At the time these data were obtained, this was the longest uninterrupted 
observational coverage of any star other than the Sun.  (MOST has since exceeded 
this duty cycle and time coverage on other targets, including Procyon: 
\citet{Procyon}.)  MOST was designed to be highly stable photometrically 
over periods of hours and self-calibrating 
(see the discussion in \citet{MOST}), 
consequently the measurements are non-differential. For some 
Primary Science Targets, fainter Secondary Science Targets in the 
adjacent direct-focus field may allow us to search for systematic 
drifts in the photometry, but no such stars were observed
with a sufficient S/N for $\kappa^1$~Ceti.

The MOST photometry is performed in a single spectral band, with a filter about 
twice as wide as the Johnson $V$ filter and slightly redward of it \citep{MOST}, 
without reference to comparison stars or flux standards.  Thus, there is 
no direct connection to any photometric system and hence, 
no absolute calibration of the MOST photometry.  
Because of this, the original data (expressed in Analog-to-Digital 
intensity units or ADU) were normalized to unity at the maximum observed 
flux of the star, occurring at JD 2,452,953.0.  
All variation amplitudes in this paper are 
also expressed in these relative units. 

To investigate rotational modulation in $\kappa^1$~Ceti, 
the 40-second exposures were binned according to the 
101.4-min orbital period of the MOST satellite, 
increasing the signal-to-noise per data point and minimizing the influence of 
orbital variations in the photometric background due to stray Earthlight.  
The electronic-version Table~\ref{tab1} lists the 
MOST orbit-averaged photometry, and those points are 
plotted in the upper panel of Figure~\ref{fig1}.

\placetable{tab1}
\placefigure{fig1}

The formal mean standard error of a single data point is 0.77 mmag (770 
parts per million or ppm), which is about twice the error 
calculated from the detected photon shot-noise. The simple average of 101 
individual 40 sec integrations would give an error of ${\sim}100$ ppm, 
which is not far from what is observed as the actual scatter of about 
200 ppm -- which includes the intrinsic variability of the star.    
Some of the photometric noise is certainly intrinsic to the star 
(e.g., granulation variations, given the star's activity)
and some is undoubtedly due to a non-Poisson instrumental 
noise resulting from the significant pointing scatter and 
the removal of the (noisy) stray light background signal
(see the next Section).

The sky background removed from the data shown in Figure~\ref{fig1} varied from 
about 21 to 25\% of the maximum flux measured from the star (on JD~2,452,953); 
the background variations are plotted in Figure~\ref{fig2}. The background
light is due primarily to light entering the instrument from the 
illuminated limb of the Earth which 
is exposed to the face of the satellite with the telescope entrance aperture.  
During commissioning, the MOST team was still investigating the background and 
exploring ways to mitigate it for future targets (e.g., 
choosing the optimum ``roll'' orientation for the satellite).  
To correct for the sky background, which is modulated with the satellite orbital 
period, but varies slightly from orbit to orbit, a running averaged 
background phased with the orbital period was subtracted from the data.  
During the $\kappa^1$~Ceti observing run, the Moon approached 
within about $14^{\circ}$ of the MOST target field,
but we observe no correlation of the sky background with the angular 
separation of the Moon and the target, or with the lunar phase 
(see the lower panels of Figure~\ref{fig2}).  
The main results of this paper are not particularly sensitive to any residual 
background variations which may be present. Although the sky 
background variability is probably the fundamental limit to the 
long-term precision of the $\kappa^1$~Ceti photometry, we note that
the 8.9-day period light photometric cycle was
highly stable in the final light curve (Section~\ref{lc-MOST}),
indicating that the background was 
correctly accounted for in the MOST data over the duration
of the $\kappa^1$~Ceti observing run. 

\placefigure{fig2}

\subsection{Nature of the Commissioning Science observations}
\label{comm}

A brief description of the ``commissioning'' observations is given
below as a record and explanation of certain limitations of 
these observations compared to the normal scientific 
operations of MOST which were initiated in January 2004, not long 
after the $\kappa^1$ Ceti data were obtained.  

The precise 
photometry that MOST was designed to obtain is based on integrating the 
signal in a large image of the telescope pupil (covering about 1500 
pixels of the CCD) projected by a Fabry microlens and field stop 
enclosing the incoming stellar beam.  In normal operations, a resolved 
sub-raster of the CCD with that pupil image is downloaded for every 
science exposure, along with Fabry images of the adjacent sky 
backgrounds, for processing on the ground.  This is known as Science 
Data Stream 2 or SDS2.  As a data backup, we also perform the pupil 
image signal integrations on board the satellite and store that 
information there for several days in case of extended loss of contact 
with our ground station network.  These data are part of Science Data 
Stream 1 (SDS1).

During commissioning, there was an unrecognized problem with a small 
fraction of the onboard memory which led to computer crashes occurring 
roughly every $1 - 2$ days if both SDS2 and engineering telemetry were 
being sent to Earth. This resulted in a fraction of
about 21\% data sent as the fully resolved SDS2 files; the rest of
our $\kappa^1$~Ceti photometry is based on SDS1 data. 
The lack of spatially resolved Fabry 
image data restricts our options of sky background and stray light 
removal.  However, our later experience (see \citet{Procyon}, for 
the example of Procyon) shows that the quality of SDS1 photometry is 
comparable to SDS2, although the latter is preferred. Even with 
our routine of downloading only SDS1 data, the onboard memory problem 
led to occasional computer crashes, which interrupted the data flow 
for one or more satellite orbits at a time, and accounts for most of 
the gaps in the light curve of Figure~\ref{fig1}.  
The small percentage of corrupt memory 
locations has since been identified, and we avoid storing data there, 
so this problem no longer affects the MOST mission.

At the  commissioning phase of the mission, we had not yet optimized the 
spacecraft pointing.  The mean pointing errors were 4.5 arcsec {\it rms\/} 
in $x$ and 6.2 arcsec in $y$, so this image wander across the Fabry microlens 
(with a field stop radius of 30 arcsec) led to larger photometric 
errors than were later obtained during normal scientific operations.  
Also, this limited our ability to center the target within the field 
stop. Jitter of the star within the field stop increases noise.  These are no 
longer serious issues for MOST.  Upgrades to the Attitude Control System 
software have led to steady improvements in pointing.  Observations of 
Procyon in January -- February 2004 were made with pointing errors of 
1.3 and 3.1 arcsec rms in $x$ and $y$ respectively; the current pointing 
performance has been improved to about 1.0 arcsec rms in both axes.

\section{SPOT VARIABILITY AND EVIDENCE FOR SURFACE 
DIFFERENTIAL ROTATION} 
\label{lc}

\subsection{Photometry of $\kappa^1$~Ceti before MOST}
\label{lc-hist}

The current explanation of the variability observed in $\kappa^1$~Ceti is that 
stellar spots cause light changes at the level of 0.02 to 0.04 magnitude. 
The changes have been observed by several ground-based observers 
to have characteristic time scales of about 9 days. 
The most extensive studies were done by \citet{Gaid2000} and 
\citet{MG2003}. From 20 years of literature data, 
\citet{Gaid2000} found several periods ranging between 9.14 and 9.46 days, each
with uncertainty at the level of 0.03 -- 0.05 days. \citet{MG2003} conducted a 
careful study over 6 seasons and determined individual periods ranging between 
9.045 and 9.406 days, with typical uncertainties of 
0.02 -- 0.05 days. The simplest explanation for such period changes 
is that spots are forming at, or 
migrating to different stellar latitudes, rotating with
different angular velocities. 

By relating the spot modulation periods to the photometric cycle, 
\citet{MG2003} suspected that $\kappa^1$~Ceti shows 
an ``anti-solar'' pattern in which the period 
of the rotation modulation tends to increase steadily 
during the activity cycle. 

\subsection{Interpretation of the MOST photometry}
\label{lc-MOST}

The photometric data discussed here consist of the orbit-averaged observations 
(binned by intervals of 101.4 min) corrected for sky background, as presented in 
Table~\ref{tab1} and the upper panel of Figure~\ref{fig1}. A double-wave
variation is clearly evident, consisting of deep minima with a depth of about 
0.04 of the maximum light, interspersed with moderately shallow depressions with 
amplitudes of about $0.005 - 0.01$. The shallow minima progressively 
move relative to the deeper ones as if two close periodicities were involved.

\placefigure{fig3}

The photometry only covers about three cycles of the dominant variation, 
so Fourier analysis is relatively ineffective at accurately identifying 
the periodic content of the variations.  We formally recover 
two significant peaks at frequencies of 0.1146 and 0.2170 day$^{-1}$, 
corresponding to periods of 8.72 and 9.22 days (actually
$4.61 = 1/2 \times 9.22$). Because of the short duration
of the observing run, these periods cannot be established to better 
than about $\pm 0.1$ day (Figure~\ref{fig3}). 
Instead of using the formal values of the derived periods, 
we use two periods, 8.9 and 9.3 days, which are derived by a
simple folding of the data in a sequential decomposition of the light curve, 
starting from the largest and most regular variation:
\begin{enumerate} 
\item The most obvious is a periodic pattern reminiscent of an eclipsing light 
curve or a spot curve, the latter due to a spot or a group of spots with 
probably well defined edges covering a moderately large fraction of the 
stellar surface. This pattern has an amplitude reaching 0.04 of the 
maximum observed flux and a period of $8.9 \pm 0.1$ days. 
An average of three cycles gives a relatively stable 
light curve, as shown in the phase diagram of Figure~\ref{fig4}.  
\item When the 8.9 day pattern is removed, another periodic variation becomes 
visible. This is shown in the middle panel of Figure~\ref{fig1}. This 
pattern is less regular than the 8.9-day variation, with the minima becoming 
progressively deeper from about 0.005 to 0.01 over the three complete cycles. 
(There is evidence that the next cycle, which is incomplete, is shallower than
the preceding one.). The three times of minima give an approximate period of 
9.3 days, where the uncertainty is harder to estimate due to the variable 
shapes of the dips. 
\item The middle panel of Figure~\ref{fig1} suggests that either the
9.3 day pattern evolved over time or there existed residual variability 
from some other source, especially at the beginning of the MOST run. The 
variability was most probably intrinsic to the star, judging by 
the progression of the depths of the 9.3 day pattern, and the 
excellent agreement between the phased minima of the 8.9 day pattern.  
We observe no correlation in the sky background variations, shown in 
Figure~\ref{fig2}, with the deviations from regular 9.3 day periodic
pattern, shown in the middle panel of Figure~\ref{fig1}; this may be
taken as indication that the sky background variations were not the
cause of these irregularities. 
\end{enumerate}

\placefigure{fig4}

\subsection{Spot modeling}
\label{lc-spots}

The observed periodic variations can be explained by 
spots formed at different stellar latitudes and carried by the rotation of the 
star with different angular velocities. With single-band photometry, little can 
be said about the type of spots, if they are hotter or cooler than the 
surrounding photosphere. However, two of the authors did develop two simple, 
entirely independent geometric spot models to try reproduce 
the observed light curve. The models differed in approach in that the first used
the minimum number of parameters and ended up describing only the larger spot 
which apparently did not evolve during the MOST observations;
only a very simplified description could be derived for the smaller spot.
The second model utilized a full set of 12 parameters required to describe two
spots and the geometry of their visibility on the star. The second model
was under-constrained, but its basic results fully confirmed the first model
in its description of the larger spot.

In the simplest terms: The larger of the two spots, with a 
rotation period of 8.9 days and a relative flux amplitude of 0.039, 
if entirely black and circular, would have
a radius of 19\% of the stellar radius 
or an angular radius from the star center of 11$^{\circ}$. 
The mean 8.9-day light curve is shown in Figure~\ref{fig4}. 
Since the light variations take place over slightly more than half of 
the rotation and the ``flat top'' is shorter than half of the
rotation, the spot appears to be situated above the equator. 
The smaller spot, causing the 9.3-day (or slightly longer)
periodicity, also if black and circular,
must be about 12\% of the stellar radius (angular radius $7^\circ$). 

Any model must face the limitation that  
the parameter without any constraint is the unperturbed level of the 
light curve, $l_0$; we arbitrarily assumed it to be at the highest 
observed level (see Figure~\ref{fig1}), but some small spots may 
have been present on the surface at all times.
While the spot longitude ($\lambda$) can be removed from the parameter
list by an appropriate adjustment of 
the phases, even the simplest model involves several adjustable
parameters: the rotation axis inclination ($i$), 
the limb darkening coefficient ($u$), the 
latitude ($\beta$), the size or the angular radius as seen from the 
center of the star ($r$), as well as the darkness ($t$) of the spot. 
The darkness ($0 \le t \le 1$) of the spot cannot be estimated from 
the single-color data, so that the minimum size of the
spot can be estimated assuming a black spot, $t=0$.  
Even such a highly simplified model, with several 
parameters fixed contains at least three fully adjustable geometrical 
parameters: $i$, $\beta$, $r$. 
The best model of the larger spot fits the primary variation to better 
than 0.001 (Figure~\ref{fig4}). The relevant parameters for the star
and the larger spot are (the symbol $\equiv$ means a fixed parameter):
\begin{eqnarray*}
P = 8.9 \pm 0.1 \,{\rm day}, \> 
i = 70^\circ \pm 4^\circ, \> 
\beta = +40^\circ \pm 7^\circ, \> 
r = 11.1^\circ \pm 0.6^\circ, \>
\lambda \equiv 0^\circ, \> t \equiv 0, \> 
u \equiv 0.8  
\end{eqnarray*}
The 8.9-day spot is 
relatively large, and may be larger if not entirely black, so that the 
assumption of it being circular -- especially in view of the clear 
evidence of the shearing differential rotation -- is highly disputable. 
Therefore, this solution should be
taken as indicative rather than of much physical significance.

The second model assumed two independent spots with different
rotation periods. It was specified by the rotational inclination $i$ of the 
star and the limb darkening parameter $u$, with all parameters describing
the two spots: the radii and flux contrasts, and
the latitude, initial longitude, and the rotation period for each spot.
Such a model required extensive tests of different combinations of
parameters. While the results are inconclusive in details
because of the large number (12) of weakly constrained parameters, 
they are fully consistent with the 
minimum-parameter model described above in the description of the first,
8.9-day spot. The second spot in the extended model appears to have 
a longer period, perhaps as long as 9.7 days, although
uncertainty of the period is large and results from the obvious
evolution of the second spot over time and a complex coupling with 
values of the remaining parameters.

The models have been unable to define accurate latitudes of both
spots, although the second, more complex model suggested a 
higher latitude for the smaller, 9.3 day spot, a situation 
which would correspond 
to the solar surface differential rotation pattern. However,
there is no question from both independent analyses that the two spots have 
unambiguously different periods, supporting the interpretation of SDR
with $\Delta \Omega/\Omega \simeq 4$~\%. This is 
a clear evidence of surface differential rotation on $\kappa^1$ Ceti,
and only the second after EK~Dra \citep{MG2003} where two
spots were also simultaneously visible. 
All other inferences were based on apparent period changes 
\citep{HB1990, HEHH1995, Gaid2000, MG2003}, interpretation
of subtle broadening effects in spectral lines,
so far detectable only in the most rapidly rotating stars
\citep{Donati1997,Donati1999} or Doppler Imaging from epoch to epoch 
\citep{CollierCameron2002}; for general reviews
of these techniques and their limitations, see 
\citet{Rice2002}, \citet{GS2004} and \citet{strass2004}.

\subsection{Searching for flare activity and transits}
\label{flares}

The MOST orbit-averaged data were also searched for short-lived variations which 
might be associated with transits of a close-in giant planet or with white-light 
flares.  The latter seemed more likely given the history of activity of 
$\kappa^1$ Ceti.  No candidates for transits or flares were found in the 
data.  We can set useful limits on such events over the month that we 
observed the star. There were no flare events or transits with 
peak-to-valley variations $>2$ mmag with durations 
longer than 200 minutes and none with variations $>3$ mmag
with durations between 2 and 200 min.

For planetary transits, this limits the size of any possible transiting bodies 
with orbital periods of about 30 days or less to approximately $0.045 - 0.055$
of the stellar radius, about half the size of Jupiter.  However, our spot 
modeling and measurements of $V \sin i$ indicate that the rotational 
inclination of the star is not close to $90^{\circ}$ so it is not 
expected that the orbital plane of any planets in the 
$\kappa^1$~Ceti system would be coincident with the line-of-sight.

\section{GROUND-BASED SPECTROSCOPY}
\label{CaII}

\subsection{The CFHT spectroscopic observations}

Several of us (D.B., S.R., E.S., G.W.) monitored the Ca II H \& K reversals of a 
number of 51~Peg systems \citep{Shk03} as well as several active, 
single stars with the Gecko spectrograph on the Canada France 
Hawaii 3.6-m Telescope (CFHT). The intention was to detect 
chromospheric activity synchronized with the planetary 
orbits. As a by-product, we determined radial velocities 
(RVs) for these stars with $\sigma_{RV}\leq$20 m s$^{-1}$ 
\citep{Wal03b}.  Here we present the various results for $\kappa^{1}$ Ceti.

The dates of the observations are listed in Table~\ref{tab2}. 
They were acquired on four observing runs in 2002 and 2003 
with the Gecko coud\'e spectrograph, fiber--fed 
from the Cassegrain focus of the CFHT. Single spectra were 
recorded at a resolution of 110,000 spanning 60 \AA\ and 
centered at 3947 \AA\ (see Figure~\ref{fig5}). The signal-to-noise was 
$\approx$ 500 per pixel in the continuum and 150 in the 
H \& K cores.  Full details 
of the data reduction are given in \citet{Shk03}.

\placetable{tab2}

\placefigure{fig5}

\subsection{Determination of $V \sin i$}
\label{vsini}

$\kappa^1$~Ceti shows moderate broadening of lines due 
to rotation at the rate of about 9 days. The broadening is
detectable in our high-resolution spectroscopy. Combined
with the rotation period and the value of the radius, 
$V \sin i$ can be used to set a constraint on the 
rotational axis inclination $i$ which is one of
free parameters in the spot models. The previous determinations
of $V \sin i$ were 3.9 km~s$^{-1}$ \citep{fek97},
4.3 km~s$^{-1}$ \citep{BM84}, 5.6 km~s$^{-1}$ \citep{sod89} 
and 3.8 km~s$^{-1}$ \citep{Mess2001}.

We determined the value of $V \sin i$ by analyzing the broadening 
of the photospheric lines in the region of
Ca~II  3933/3968 \AA\ in our CFHT spectra.
We used the broadening function formalism \citep{Ruc99,Ruc02}
of the deconvolution of photospheric spectra of rotating
stars with spectra of sharp-line, slowly-rotating
stars. The approach leads to removal of all common agents 
of line broadening, not only of the instrumental component, 
but also of the tangential turbulence and of all
other components which are the same for the program 
and the template stars.
For $\kappa^1$~Ceti, we used $\tau$~Ceti which is a slightly
cooler dwarf, G8V, and rotates very slowly, 
$V \sin i = 0.6$ km~s$^{-1}$ \citep{fek97}.

\placefigure{fig6}

For determination of the broadening function, the spectra must be rectified 
and normalized. As shown in Figure~\ref{fig5},
the rich photospheric spectrum in both $\kappa^1$~Ceti and $\tau$~Ceti
is superimposed on strongly variable far wings of the Ca~II absorption.
We removed sections of the spectrum where the
wings fall below 1/3 of the maximum flux and applied
a simple, upper-envelope rectification 
between high points of the quasi-continuum. 
Figure~\ref{fig6} shows the broadening function for one 
of the CFHT spectra, obtained on August 26, 2003
(see Table~\ref{tab2}). 
Instead of least-squares fits of the limb-darkened 
rotational profile \citep{gray1992} to the broadening functions
(with the limb darkening coefficient assumed
$u=1.0$, appropriate for the UV in late-type stars),  we
matched the half-widths at half maximum; such
$V \sin i$ determinations are less sensitive to the
residual uncertainties at the base of the broadening function.
Our average $V \sin i = 4.64 \pm 0.11$ km~s$^{-1}$ is
compatible with previous determinations. We note that \citet{Gaid2000}, 
using an almost identical value of $V \sin i = 4.5$ km~s$^{-1}$, 
interpreted it as implying $40^\circ < i < 90^\circ$. 
The main spot in both model fits is located 
at moderate stellar latitude above the equator.  
If the period of $8.9 \pm 0.1$ days were ascribed to the stellar 
equator, then the best fitting inclination of 
$i = 60^{\circ} \pm 5^{\circ}$ and the measured $V \sin i$ combine 
to give a stellar radius of $R \simeq 0.95 \pm 0.10 \, R_{\odot}$, 
which is certainly reasonable for a G5V star.

\subsection{The K-line residuals}
\label{K-res}

The 7 \AA\ region centered on the K-line which was used in the analysis is
shown for a single spectrum in Figure~\ref{fig7} where spectral intensity is
normalized to unity at the wavelength limits. In Figure~\ref{fig8} the
residuals from the average normalized spectra are shown for each night
(where pairs of spectra on the same night have been averaged). Activity
is confined entirely to the reversals, within about 0.94 \AA\
around the center of the emission core or within $\pm 35.8$ km~s$^{-1}$. If
this broadening is caused by the velocity field, it is some 7.7 times larger
than $V \sin i$. To visualize the changes in the reversals,
the integrated flux residuals shown in Figure~\ref{fig9} are expressed in \AA\ 
relative to the restricted continuum height shown in Figure~\ref{fig7}.
This height corresponds to about 1/3 of the local photospheric
continuum around the Ca~II lines so that the integrated residuals
are proportional, but not identical, to the emission equivalent width. They
are listed in column 4 of Table~\ref{tab2}.

\placefigure{fig7}
\placefigure{fig8}

The best fitting periodic variation for the K residuals 
in Table~\ref{tab2} over the four observing runs in 2002 and 2003
has a period of $9.332 \pm 0.035$ days.  The residuals are plotted
as a function of phase in Figure~\ref{fig9} with the arbitrary
initial epoch of the minimum residual flux at 
$T_0(JD_{hel}) = 2,452,485.117$ (the minimum flux epoch 
can be estimated to $\pm 0.1$ days). The data from 2002 are
shown as open symbols and those from September 2003 are solid. Although
the data from 2002 cover only 3 rotations, with the data from 2003 the
span in the rotations is extended to 44.

\placefigure{fig9}

Since the period is largely defined
by the 2002 observations, Figure~\ref{fig9} 
implies that the rotation associated with the enhanced Ca~II
activity has remained quite stable. 
The 9.332 day period is entirely compatible
with that found by \citet{Bal83} ($9.4 \pm 0.1$ day) suggesting that the
enhanced activity has in fact been stable at the same latitude for at
least 36 years (over 1400 rotations). This period
is also compatible with the 9.3 day period of the less
prominent spot that we see in the MOST 
observations (see Section~\ref{lc-MOST}).  
Given this apparent stability, it
is reasonable to extrapolate variations in the K-line residuals to the
epoch of the MOST photometry which corresponds to only 7
additional rotations. The bottom panel of Figure~\ref{fig1} shows 
the K-line flux changes which can be related to
the photometric variations due to
the 9.3 day spot seen in the MOST photometry. The maxima of the 
two curves coincide. The interpretation of this coincidence 
is not obvious as the enhanced Ca~II could be expected to 
accompany the dark spot, unless what we see as the 9.3 day 
photometric signal is actually a bright spot and the minima are due
to disappearance of the bright plage region.

\subsection{The Radial Velocities}
\label{RV}

Radial velocities were estimated from $\approx$ 20 \AA\/ of the photospheric
spectrum ($3942 - 3963$ \AA, see Figure~\ref{fig5})
with the ``fxcor'' routine in
IRAF\footnote{IRAF is distributed by the National Optical Astronomy
Observatories, which is operated by the Association of Universities for
Research in Astronomy, Inc. (AURA) under cooperative agreement with the
National Science Foundation.}, with Th/Ar comparison lines providing
dispersion correction. More detail is given in \citet{Wal03b}. The first
spectrum of the series in 2002 acted as the template from which differential
radial velocities ($\Delta$RV) were measured for all the spectra. The
$\Delta$RV values are listed in Table \ref{tab2}. They
are plotted against the 9.332 day rotational phase determined 
from the K-line residuals in Figure~\ref{fig9}
where the open symbols correspond to data from 2002 and the solid
points to September 2003. There is a remarkable
consistency between the $\Delta$RVs from both years suggesting that the
underlying $\Delta$RV technique is highly stable even without an
imposed fiducial (such as iodine vapor).

From the 2002 data we derive $\sigma_{RV}= $21.8 m~s$^{-1}$ and 
23.6 m~s$^{-1}$ when combined with 2003. These values are very similar to
$\sigma_{RV}= $24.4 m~s$^{-1}$, found over 11 years by 
\citet{Cumming1999}. For comparison, a planet which 
induces a reflex radial velocity variation $<50$ m~s$^{-1}$ and is
tidally synchronized to the star, would have $M \sin i \simeq 0.74 M_J$
and a = 0.084 AU. Consequently, $\kappa^1$~Ceti could still harbor a giant
planet in a tidally locked orbit which we would not have detected.

In 2003 the $\Delta$RVs appear significantly different between the two
nights, something which seems to be reinforced by the 
consistency within the pairs of $\Delta$RVs. While
a planetary perturbation cannot be ruled out by the 2002 data and other 
Precision Radial Velocity 
studies, the difference in 2003 might be associated with the velocity
field of the star itself. The increase of velocity with increasing
K-line strength is consistent with the extreme event in 1988.5 seen
by \citet{Wal95}. However, the data are insufficient to
discuss the point further here.

\section{ABSENCE OF P-MODE OSCILLATIONS }
\label{p-mode}

In Figure~\ref{fig10}, we present the Fourier amplitude spectrum of the complete 
set of unbinned photometry (at full temporal resolution of about one minute), 
in which the lower frequency modulations have been 
filtered from the data.  Also, the known satellite orbital frequency and its 
harmonics have been removed from the data, to avoid any confusion with residual 
stray light variations.
Although the spectrum is plotted only to a frequency of 4 mHz 
(period = 4.17 min), 
the noise is level out to the Nyquist frequency of about 8 mHz, and there 
are no significant periodic signals detected in that range.  The white dashed 
line represents the theoretical Poisson noise for the data.  

\placefigure{fig10} 

Although there are a few peaks visible at frequencies below 1 mHz, there is no 
evidence for the pattern of nearly equal spacing associated with 
p-mode oscillations of low degree and high overtone (as are seen in the 
Sun).  There are several isolated peaks in the range 0.1 -- 1.0 mHz 
(periods of 2.8 hr to 17 min) which may be real but are too few and 
widely separated to attempt a meaningful fit to a model.  

At a detection level of $3\sigma$, we exclude coherent oscillations larger than 
about 27 ppm around 1 mHz, 26 ppm at 2 mHz, 
and 24 ppm at 3 mHz. (The last is in the frequency range of the 
solar oscillations.)  Thus, we would not expect to detect
p-mode oscillations in $\kappa^1$~Ceti if they were at the level observed in 
photometry of integrated light of the Sun; i.e., about 4 -- 10 ppm 
\citep{frohlich1997}.

However, knowledge of the nature of p-mode oscillations in other stars is very 
limited, and the first space-based photometry of Procyon \citet{Procyon} 
obtained by MOST yielded an unexpected null result despite a detection limit 
at least twice as good as in these commissioning data of $\kappa^1$~Ceti.
We are only beginning to explore the characteristics of p-mode oscillations in 
luminosity for stars beyond the Sun.  The moderately young age of $\kappa^1$ 
Ceti, its relatively rapid rotation compared to the Sun, and its 
later-than-solar 
spectral type may be contributing factors to the absence of large-amplitude 
(i.e.,$> 20$ ppm) p-modes in this star.

\section{CONCLUSIONS}
\label{concl}

We have presented data on the photometric and spectroscopic variability of the 
moderately young, solar-type star $\kappa^1$~Ceti. We describe our continuous
photometric observations over one month in November 2003 using the MOST satellite 
and our high-resolution spectroscopic CFHT observations over four observing runs
in 2002 and 2003. The MOST data (averaged over the 101.4-min orbital period of
the satellite) have an estimated point-to-point accuracy of $\sim 0.2$ 
millimagnitude or $\sim 200$ parts per million, and an unprecedented temporal 
coverage of 30.5 days with a completeness of 96\%.

The MOST orbit-averaged photometry of $\kappa^1$~Ceti shows two distinctly 
different periodicities -- a well-defined 8.9-day variation and a smaller, less 
regular variation whose period appears to be around 9.3 days or slightly 
longer.  The best interpretation of these variations is that two spots (or 
compact spot groups) are at different latitudes on the stellar surface, moving
at angular rates which are different by $\Delta\Omega / \Omega \simeq 4$~\%.
This is one of the most direct evidences of surface
differential rotation in a cool star other than the Sun and 
the second such after EK~Dra \citep{MG2003}. All other
measurements of SDR in other stars have been more indirect, either through 
season-to-season variations in spot 
period variability or through subtle broadening effects in spectral line profiles, 
so far observable only in the most rapidly rotating stars. 

The details of the smaller slower moving spot on $\kappa^1$~Ceti depend on how 
one describes and removes the larger spot, 8.9 day, variability and on the
treatment of the slowly varying residuals. The multi-parametric modeling
of the spots suggests that the smaller spots is located at a higher latitude,
which would suggest a solar-like SDR pattern. However, 
independent spectroscopy of the star, some of which was obtained only a few 
variation cycles before the MOST photometry, reveals that maxima of the Ca~II 
K-line emission reversals are synchronized with a period of about 9.3 days, 
lending support to that period identification for the second spot. While showing
significant evolution in size, shape and/or flux contrast even during our month 
of observations, the 9.3-day spot may in fact be the more stable one and perhaps 
visible over decades, based on earlier observations of $\kappa^1$~Ceti in the
literature.   
The shorter, 8.9 day, periodicity suggests that $\kappa^1$~Ceti 
rotates faster than previously supposed and thus may be even younger, 
based on the correlation between equatorial rotation period
and age in solar-type stars.  

The unbinned MOST photometry, with an integration time of 40 seconds and a 
sampling of once per minute, sets limits on (1) the presence of 
p-mode oscillations, 
(2) the eruptions of flares over one month of monitoring of this historically 
active star, and (3) the occurrence of planetary transits in the system.  The 
data effectively rule out p-mode oscillations larger than 25 parts per million 
(ppm) at the 3--$\sigma$ limit.  There were no flares with durations from 
2 to 200 min whose white-light amplitudes could be larger than 2 -- 3 mmag.  
There can be no planets with orbital periods shorter than 30 days 
larger than about 0.5 Jupiter diameter 
in an orbit plane which would lead to transits as seen from Earth. 
Within the value of $\sigma_{RV}\sim $23 m~s$^{-1}$ 
we determine, there could be a close Jupiter-mass companion in an inclined orbit.  

The information gleaned from this early set of MOST data, obtained before the 
satellite performance was optimized for normal operations on a target which was 
not part of the original planned scientific program, indicates the tremendous 
power of precise, nearly continuous photometry spanning many weeks, which is 
possible from a dedicated instrument in the right orbit in space.  There are many
more applications like the ones described in this paper, and instruments like 
MOST can be an effective tool in studying the surface properties of rapidly 
rotating solar-type stars.

\acknowledgements

We thank the anonymous referee for excellent suggestions which helped
us to improve the quality and consistency of the paper.

The MOST project is funded and supervised by the Canadian Space Agency.  
We are grateful to the engineers at Dynacon Inc., the University of Toronto
Institute for Aerospace Studies -- Space Flight Laboratory, the University of
British Columbia Department of Physics and Astronomy, and the University 
of Vienna and Austrian Technical University for their contributions to the 
success of the MOST mission. D.B.G., J.M.M., A.F.J.M., S.M.R., E.S., 
G.A.H.W. were supported by the Natural Sciences and Engineering Research 
Council Canada. A.F.J.M. is also supported by FCAR (Qu\'ebec). 
W.W.W. received support from the Austrian Space Agency 
and the Austrian Science Fund (P14984).

This research has made use of the SIMBAD database, operated at the CDS,
Strasbourg, France and accessible through the Canadian
Astronomy Data Centre, which is operated by the Herzberg Institute of
Astrophysics, National Research Council of Canada.

\noindent
\clearpage               

\figcaption[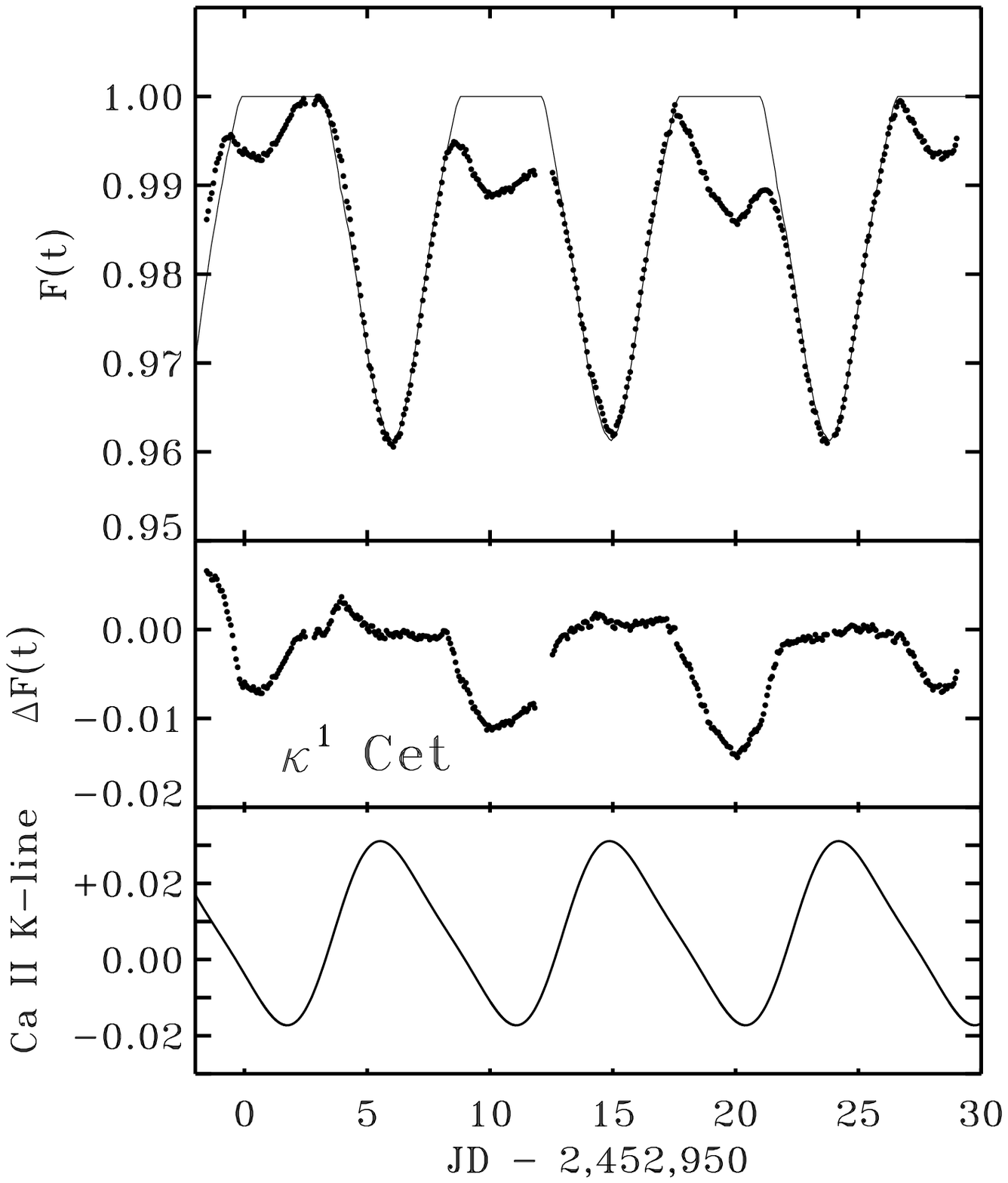] {\label{fig1}
({\em upper panel}) The light curve of $\kappa^1$~Cet obtained from 30.5 
days of continuous observations with MOST.  The points are averages over 
the 101.4 min orbital period of the satellite. The variable background from 
Earthlight has been subtracted.  The data are presented as relative
signal normalized to the maximum flux observed during the run. The solid 
curve is the best fit of a simple single spot model to the dominant 8.9-day 
variation in the data (see Section~\ref{lc-spots} and Figure~\ref{fig4}). 
({\em middle panel}) The residual variation after subtraction of the 
8.9-day model 
curve plotted in the upper panel shows clear variations whose minima are 
spaced by about 9.3 days (see Section~\ref{lc-MOST}).  ({\em bottom panel})  
The predicted variation of the Ca~II K-line emission flux based on the 
spectroscopic observations obtained only two months before 
the MOST run (see Section~\ref{K-res} and Figure~\ref{fig9}).
} 

\figcaption[Kap1Cet_fig2.ps] {\label{fig2}
({\em upper panel}) The variable sky background measured during our 
observations, expressed as a fraction of the maximum stellar signal on
JD~2,452,953. This background is primarily due to stray light entering the
instrument focal plane from the illuminated limb of the Earth.
({\em lower panels}) The angular separation of the Moon from the star 
and the corresponding lunar phase during the MOST observations.  There is
no evident correlation with the sky background in the upper panel.
}

\figcaption[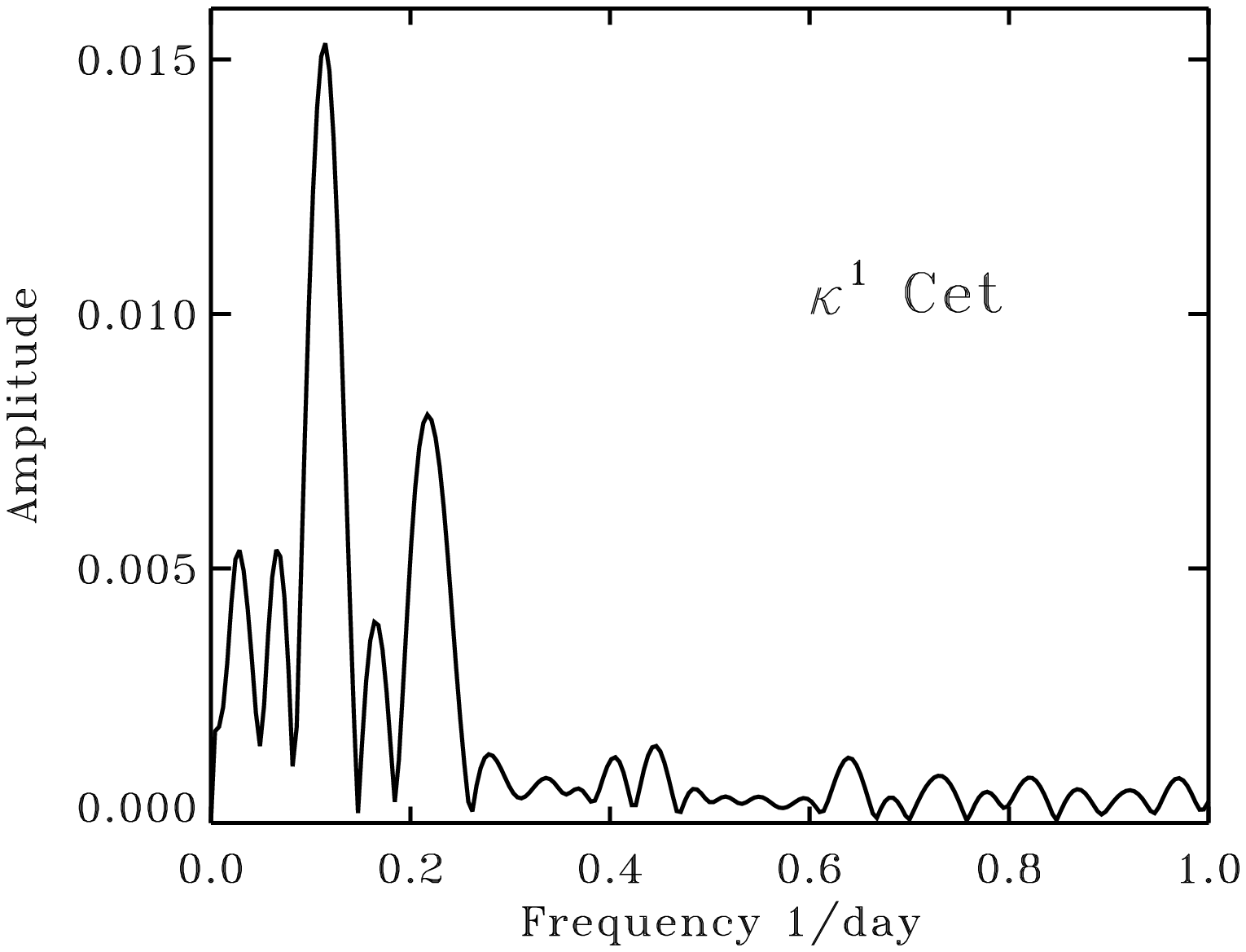] {\label{fig3}
The periodogram based on the original data (the points in
the upper panel of Figure~\ref{fig1}). The periodogram does not 
show any significant frequencies above one cycle per day.
}

\figcaption[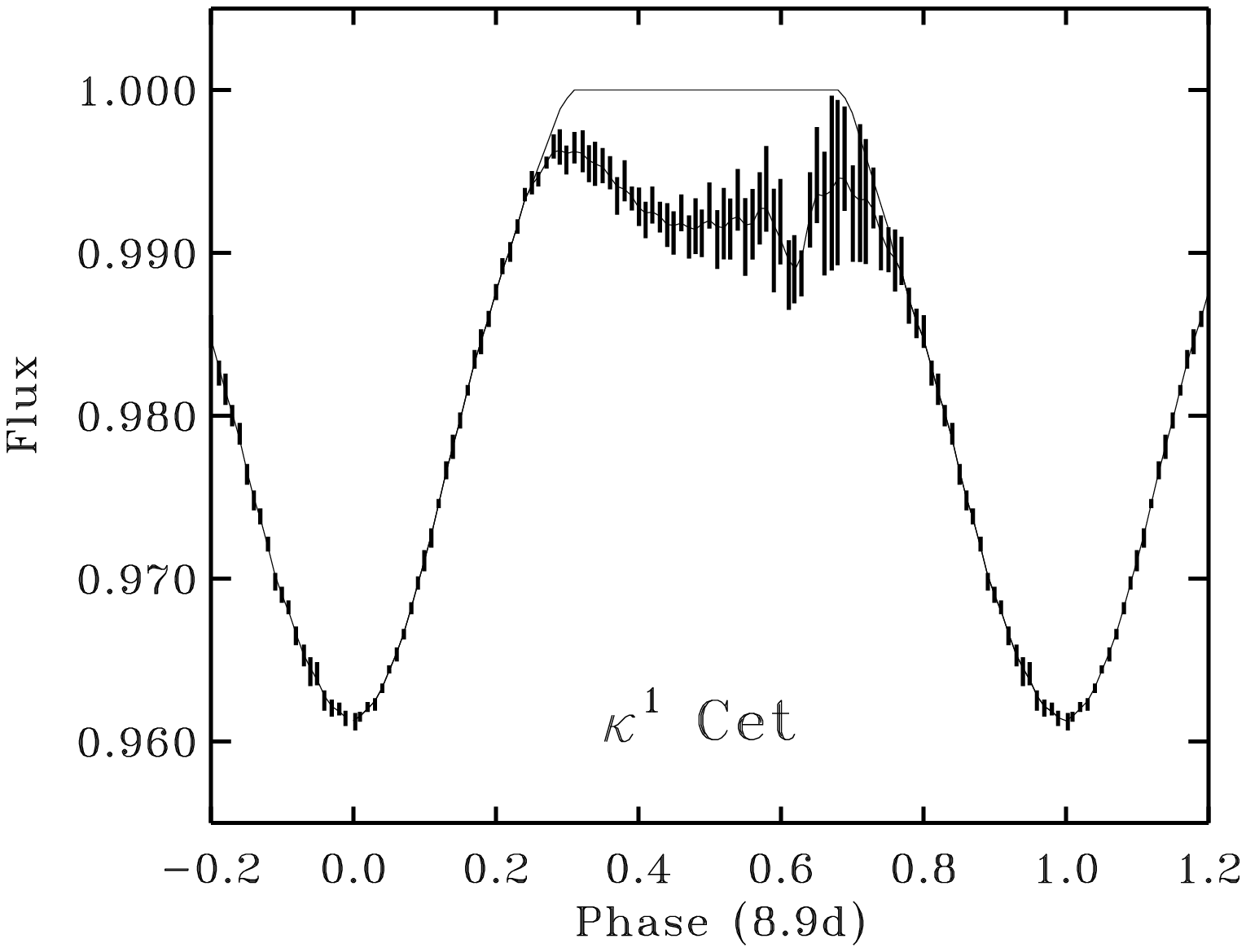] {\label{fig4}
The data in the upper panel of Figure~\ref{fig1} are shown
phased to a period of 8.90 days.
The error bars show the $\pm 1$--$\sigma$ scatter in the phase-binned data. 
The scatter is larger at phases when the second-spot variability is 
best noticeable, consistent with visual inspection of the light curve. The 
single spot model from the upper panel of Figure~\ref{fig1} 
is indicated by the thin solid line.
}

\figcaption[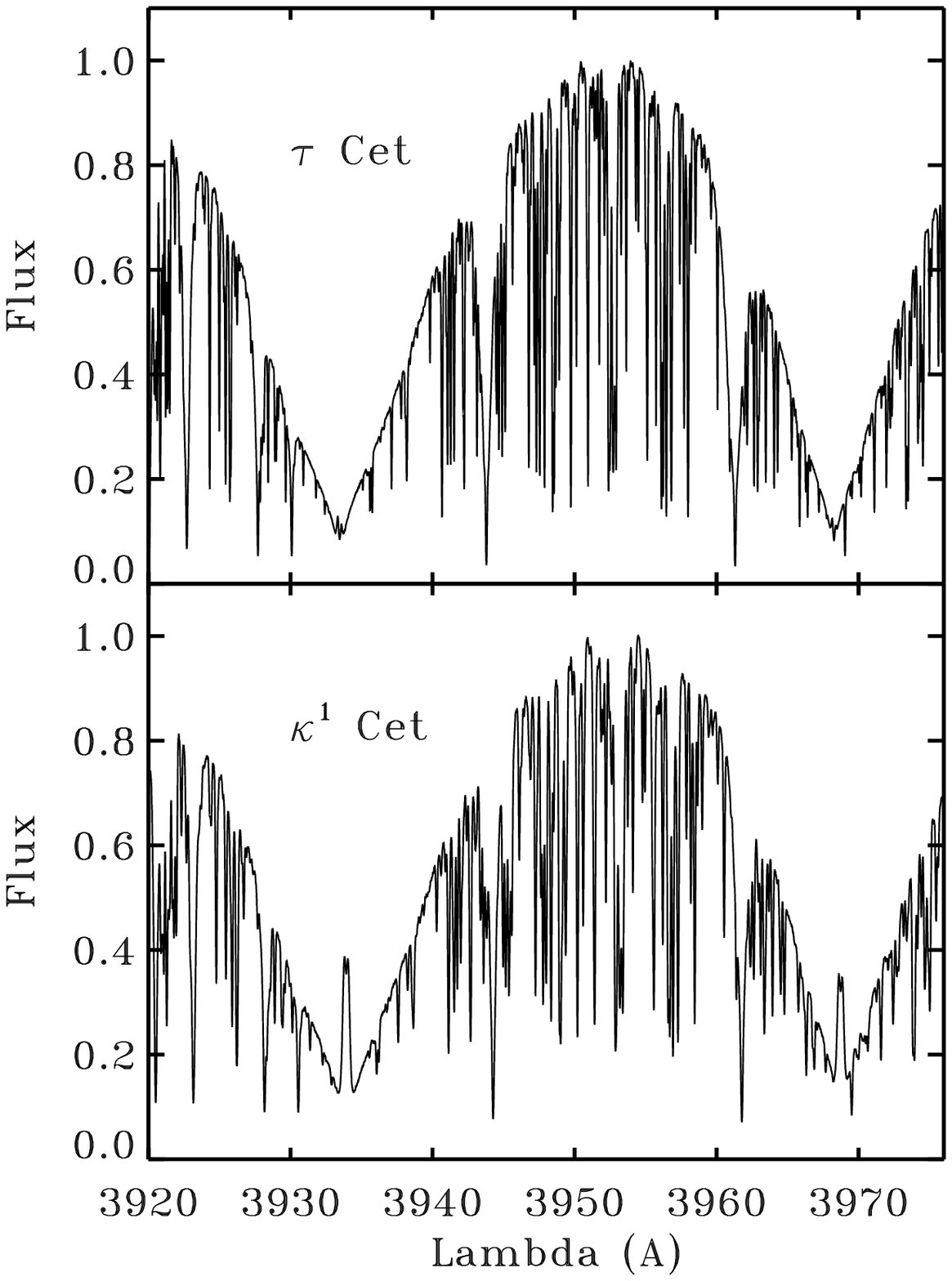] {\label{fig5}
Typical CFHT spectra of $\kappa^1$~Ceti and of $\tau$~Ceti
expressed as normalized flux versus wavelength; the latter
star served as the sharp line template.  The broadening functions 
for $\kappa^1$~Ceti were determined from the upper parts of the
photospheric spectra after removal from the spectra of 
the lower parts close to the cores at $<1/3$ of the maximum height.
}

\figcaption[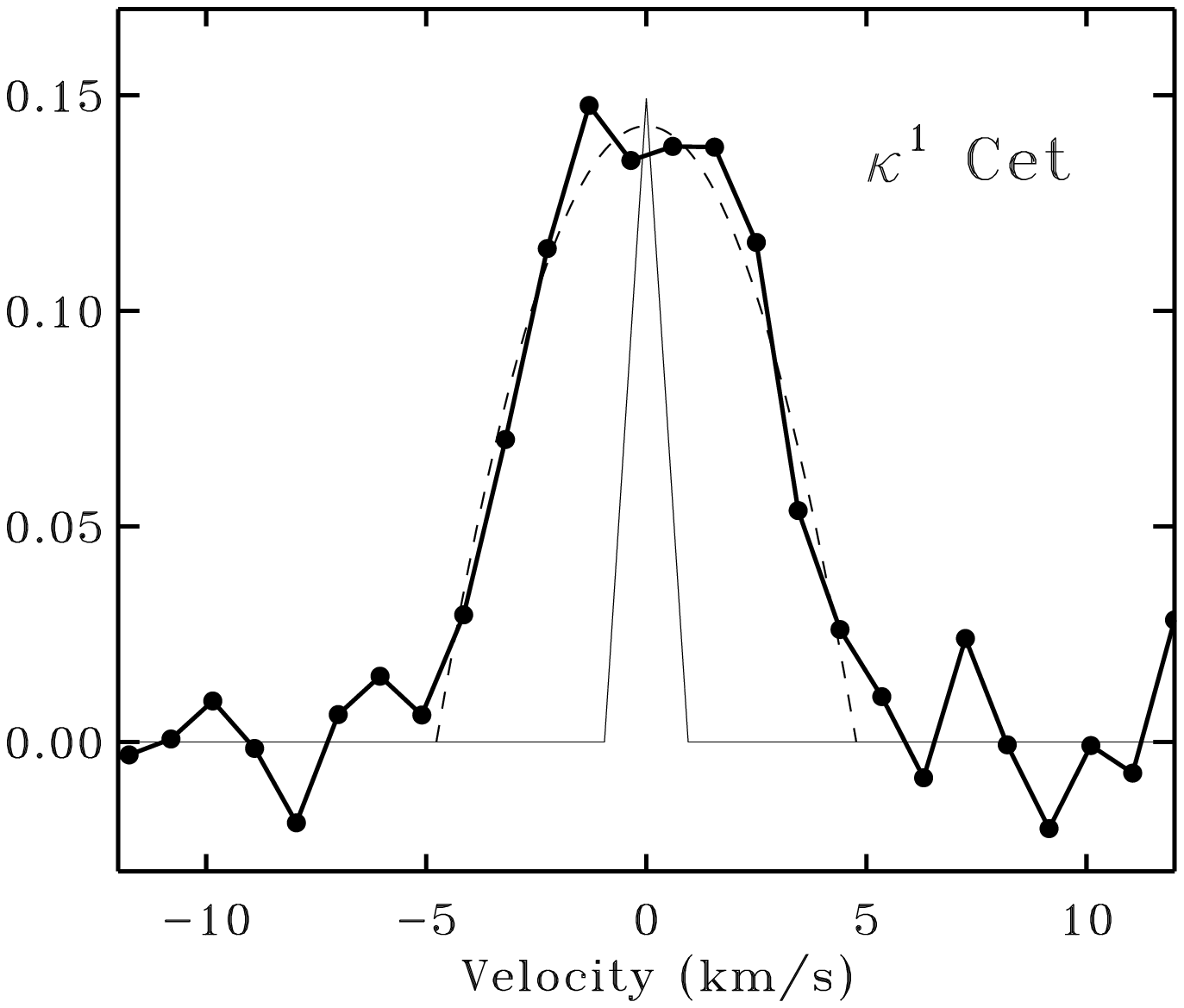] {\label{fig6}
The broadening function for $\kappa^1$~Ceti determined relative 
to $\tau$~Ceti for one of the CFHT spectra. The broadening function gives a 
one-dimensional image of the rotating star, projected into the radial 
velocity space; it can be used for determination of the value of 
$V \sin i$. The broadening function is dimensionless and -- for perfect
match of the spectral types -- its integral equals unity.
The thin line shows the intrinsic spectral resolution of 
the method while the broken line gives the calculated rotational profile 
for $V \sin i = 4.64$ km~s$^{-1}$ and the limb darkening of $u=1.0$.
The spike in the center shows the intrinsic resolution of the technique.
}

\figcaption[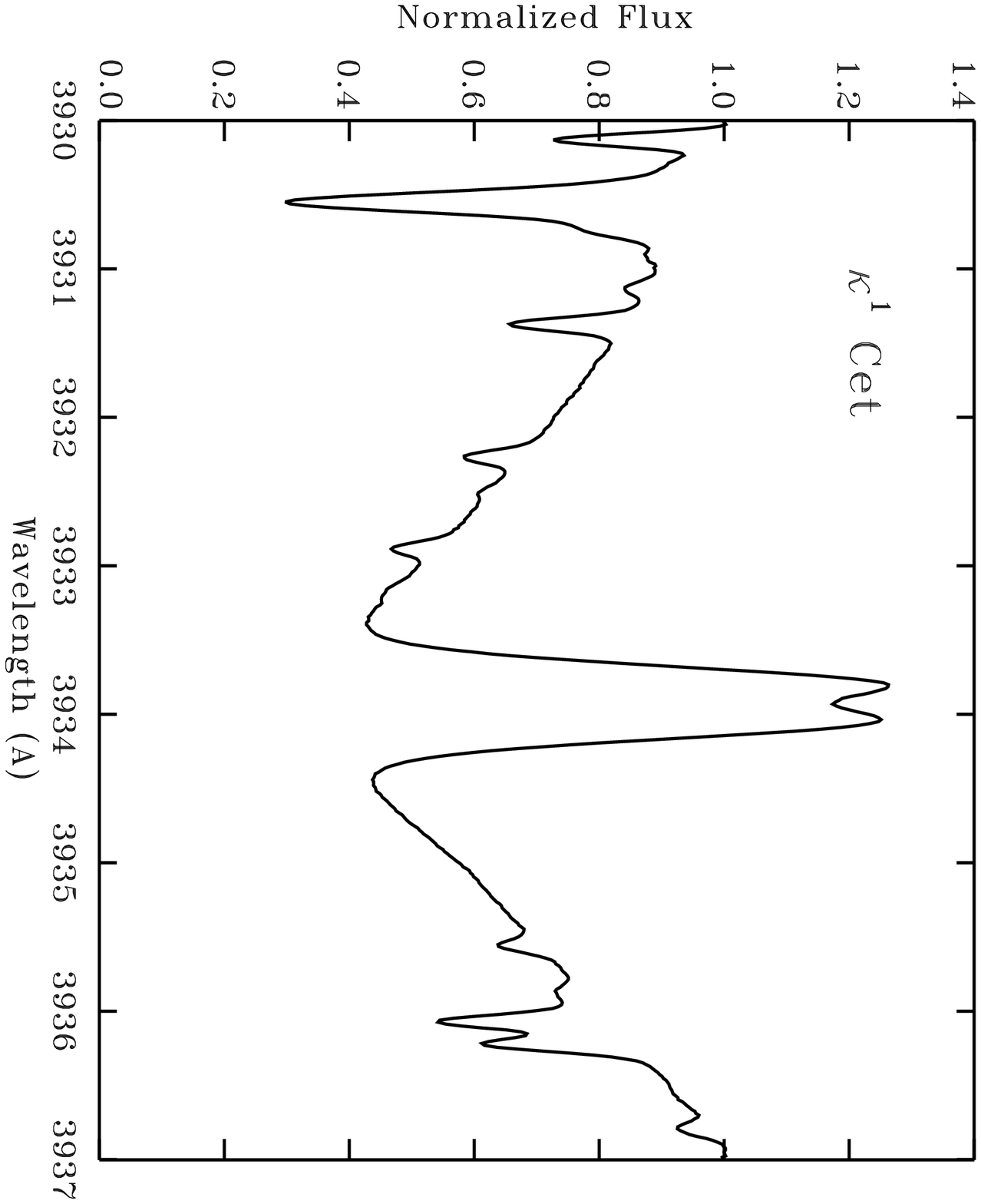] {\label{fig7}
The Ca~II K-line emission reversal, normalized at the ends of the 
spectral window in a single typical CFHT spectrum.
}

\figcaption[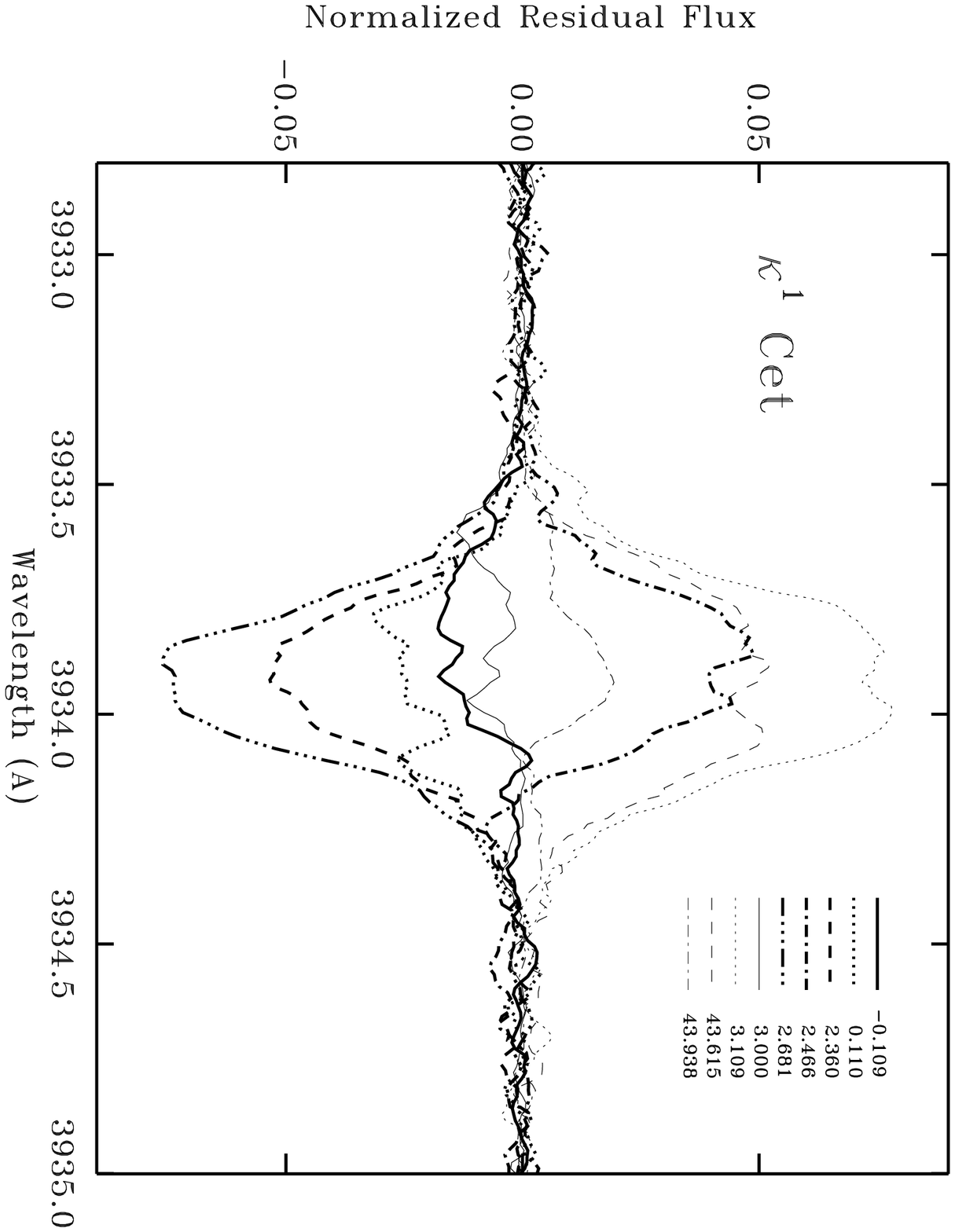] {\label{fig8}
Residuals from the Ca~II K-line mean averaged normalized profile, as 
shown in Figure~\ref{fig7}, for individual nights, with
one to three observations averaged if close in time. Note that all 
changes are confined strictly to the Ca~II emission reversal. The 
phases of the 9.332-day periodicity are given in the legend; they 
correspond to those in Table~\ref{tab2} where individual observations
are listed.
}

\figcaption[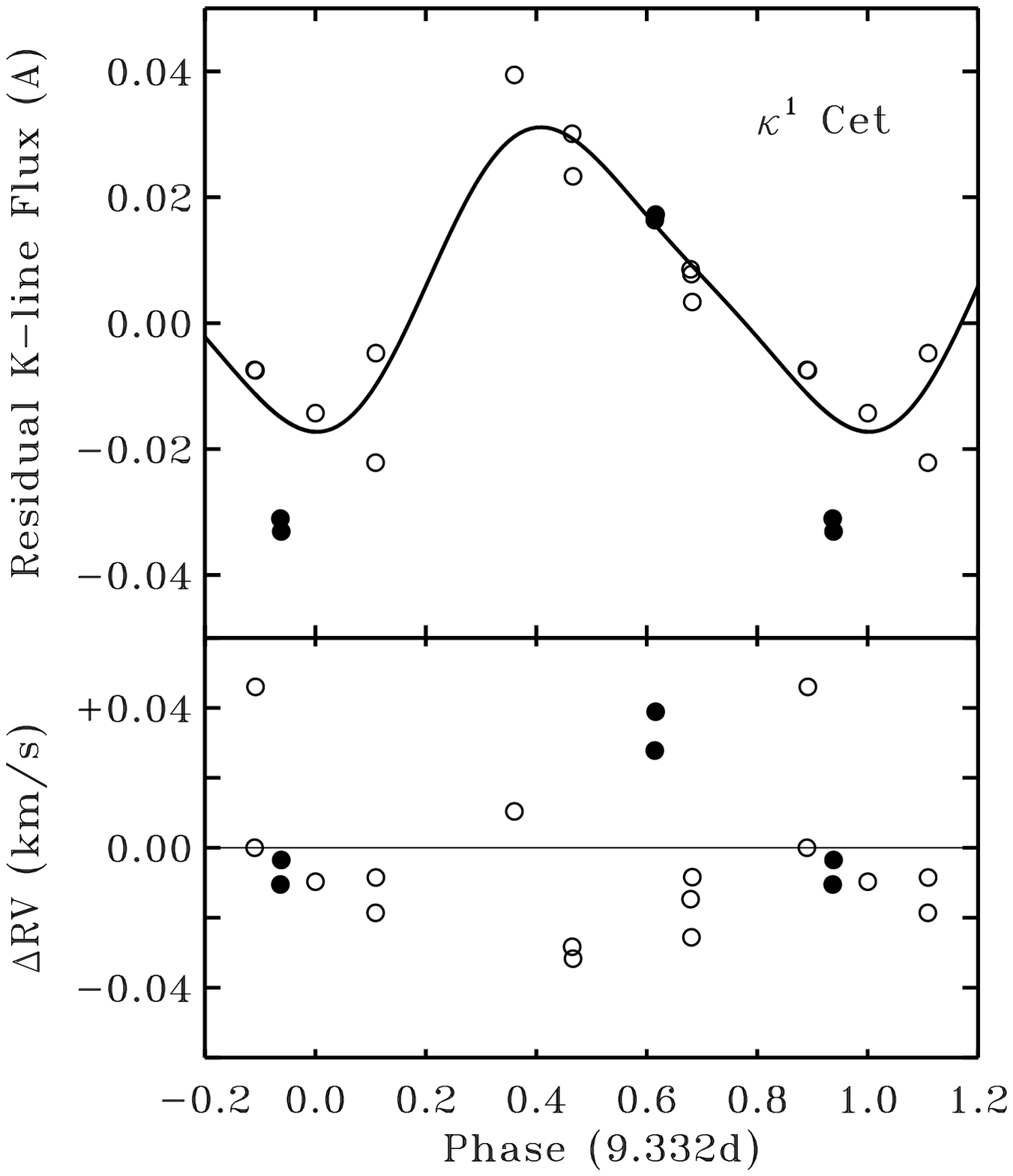] {\label{fig9}
({\em upper panel}) 
Integrated residuals of the Ca~II K-line emission expressed in \AA\ 
and plotted versus the 9.332-day rotation phase. 
Data points from 2002 are marked by open circles while those 
from 2003 are solid circles.  
({\em lower panel}) Radial velocity deviations from the 
first spectrum. 
}

\figcaption[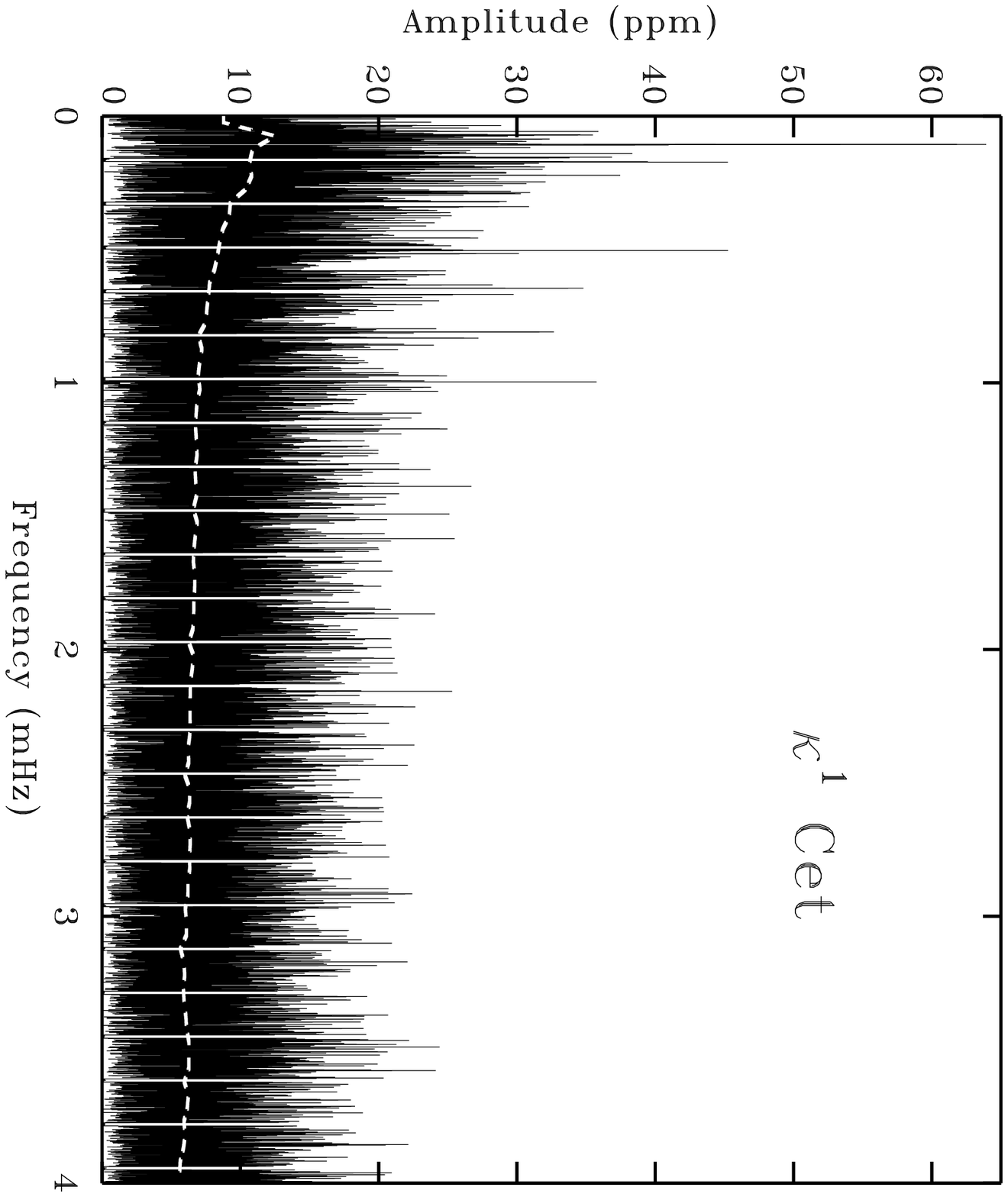] {\label{fig10}
The Fourier amplitude spectrum of the unbinned MOST photometry with 
40-sec integrations and one-minute sampling interval.  The maximum 
frequency in this plot (4 mHz) corresponds to a period of 250 
seconds; there is no noticeable change in noise level nor any signals
evident at frequencies up to the Nyquist frequency of about 8 mHz.
The amplitudes for frequencies corresponding to the satellite orbital
frequency and its harmonics have been removed to avoid any contamination
by orbital modulation of the stray light background.  The dashed line
is the formal 1--$\sigma$ noise level in the spectrum. 
} 

\clearpage                    

\begin{deluxetable}{ccc}
\tabletypesize{\footnotesize}
\tablecaption{$\kappa^{1}$ Ceti: 
MOST photometric data, orbital averages   
\label{tab1}
}
\tablewidth{0pt}
\tablehead{   
\colhead{JD(hel)-2,452,948\tablenotemark{a}} &
\colhead{Normalized flux\tablenotemark{b}} &
\colhead{Background flux\tablenotemark{c}}
}
\startdata
    0.461914  &  0.986152  &  0.213726 \\
    0.525269  &  0.987092  &  0.209837 \\
    0.598022  &  0.988443  &  0.213698 \\
    0.668823  &  0.989125  &  0.212024 \\
    0.736816  &  0.990238  &  0.210084 \\
\enddata
\tablenotetext{a}{The Julian Date at the effective center 
of a MOST orbit of 101 minutes.}
\tablenotetext{b}{The observed flux normalized to the maximum on
JD 2,452,953.0.}
\tablenotetext{c}{The background flux normalized to the maximum 
of the stellar signal on JD 2,452,953.0.}
\tablecomments{The first five lines of the full table are
shown here. The table is available only as a computer ascii file.} 
\end{deluxetable}

\clearpage

\begin{deluxetable}{ccccccl}
\tabletypesize{\footnotesize}
\tablecaption{$\kappa^{1}$ Ceti: 
CFHT Radial velocities and Ca II K line residuals
\label{tab2}
}
\tablewidth{0pt}
\tablehead{   
\colhead{UT Date}&
\colhead{JD\tablenotemark{a}} & 
\colhead{$\phi_{rot}$\tablenotemark{b}} &
\colhead{Residual K\tablenotemark{c}} &
\colhead{$\Delta$RVs}&
\\

\colhead{} &
\colhead{} &
\colhead{} &
\colhead{Flux ($\pm$ 0.0017 \AA)} &
\colhead{km s$^{-1}$}&
}
\startdata
7/28/2002 & 2452484.0907 &     \phn $-$0.110 &  $-$0.0074 & \phs 0.0000\\
7/28/2002 & 2452484.1062 &     \phn $-$0.108 &  $-$0.0075 & \phs 0.0460\\
7/30/2002 & 2452486.1395 &\phs \phn  0.110   &  $-$0.0048 &   $-$0.0085\\
8/20/2002 & 2452507.1459 &\phs \phn  2.360   &\phs 0.0394 & \phs 0.0104\\
8/21/2002 & 2452508.1227 &\phs \phn  2.465   &\phs 0.0301 &   $-$0.0283\\
8/21/2002 & 2452508.1371 &\phs \phn  2.467   &\phs 0.0233 &   $-$0.0317\\
8/23/2002 & 2452510.1235 &\phs \phn  2.680   &\phs 0.0085 &   $-$0.0147\\
8/23/2002 & 2452510.1379 &\phs \phn  2.681   &\phs 0.0078 &   $-$0.0256\\
8/23/2002 & 2452510.1524 &\phs \phn  2.683   &\phs 0.0034 &   $-$0.0084\\
8/26/2002 & 2452513.1172 &\phs \phn  3.000   &  $-$0.0143 &   $-$0.0097\\
8/27/2002 & 2452514.1330 &\phs \phn  3.109   &  $-$0.0222 &   $-$0.0186\\
9/09/2003 & 2452892.1321 &\phs      43.615   &\phs 0.0164 &\phs  0.0278\\
9/09/2003 & 2452892.1465 &\phs      43.616   &\phs 0.0172 &\phs  0.0389\\
9/12/2003 & 2452895.1384 &\phs      43.937   &  $-$0.0311 &   $-$0.0105\\
9/12/2003 & 2452895.1529 &\phs      43.939   &  $-$0.0331 &   $-$0.0035\\
\enddata

\tablenotetext{a}{Julian Date at mid-exposure time. Exposure times are 1200 s.}
\tablenotetext{b}{Rotational phases calculated using $P_{rot}$ = 9.332 days,
$T_0(JD_{hel}) = 2,452,485.117$. }
\tablenotetext{c}{Integrated residual Ca II K emission flux (in \AA) from a
mean of all spectra (see Figure 8 and the text for details).}
\end{deluxetable}

\end{document}